\documentclass[twocolumn,showpacs,preprintnumbers,amsmath,amssymb,floatfix,prb]{revtex4}

\usepackage{graphicx}
\usepackage{dcolumn}
\usepackage{bm}

\begin{document}

\title{
Cation diffusion and hybridization effects at the Mn-GaSe(0001)
interface probed by soft X-ray electron spectroscopies}

\author{S. Dash$^{1}$, G. Drera$^{1}$, E. Magnano$^{2}$, F. Bondino$^{2}$,
P. Galinetto$^{3}$, M.C. Mozzati$^{3}$,  G. Salvinelli$^{1}$, V.
Aguekian$^{4}$, and L. Sangaletti$^{1}$\\}

\affiliation{%
$^{1}  $Interdisciplinary Laboratories for Advanced Materials
Physics (I-Lamp) and Dipartimento di Matematica e Fisica,
Universit\`{a}
Cattolica, via dei Musei 41, 25121 Brescia (Italy)}%

\affiliation{$^{2}$ IOM-CNR, Laboratorio TASC, S.S. 14, Km 163,5 I-34149 Basovizza (Italy)}%

\affiliation{ $^{3}$CNISM and Dipartimento di Fisica, Universit\`{a} di Pavia, Via Bassi 6 (Italy)}%

\affiliation{$^{4}$ Solid State Physics Department, V. Fock,
Institute of Physics, Saint-Petersburg State University,
Petrodvoretz, 198904 S.-Petersburg, Russia}

\date{\today}

\begin{abstract}
The electronic properties of the Mn:GaSe interface, produced by
evaporating Mn at room temperature on a $\epsilon$-GaSe(0001) single
crystal surface, have been studied by soft X-ray spectroscopies.
Substitutional effects of Mn replacing Ga cations and Mn-Se
hybridization effects are found both in core level and valence band
photoemission spectra.  The Mn cation valence state is probed by XAS
measurements at the Mn L-edge, which indicate that Mn diffuses into
the lattice as a Mn$^{2+}$ cation with negligible crystal field
effects. The Mn spectral weight in the valence band is probed by
resonant photoemission spectroscopy at the Mn L-edge, which also
allowed an estimation of the charge transfer and Mott-Hubbard
energies on the basis of impurity-cluster configuration-interaction
model of the photoemission process. The charge transfer energy is
found to scale with the energy gap of the system. Competing effects
of Mn segregation on the surface have been identified, and the
transition from the Mn diffusion through the surface to the
segregation of metallic layers on the surface has been tracked by
core-level photoemission.
\end{abstract}

\pacs{Valid PACS appear here}
\maketitle

\section{Introduction}

The III-VI semiconductors GaSe, InSe, GaTe, and GaS have received
considerable interest in the last few years because they show
remarkable nonlinear optical properties and they are regarded as
promising materials for photo-electronic applications
\cite{010,020,030, 001, 002}, even in the form of
nanowires\cite{004}. In the particular case of GaSe and GaS, the
interest on these systems has been recently renewed due to the
possibility to obtain ultrathin layer transistors based on
atomic-thin sheets \cite{003}.

The magnetic properties of these systems doped with transition metal
ions (e.g. Mn\cite{pekarek}, or Fe\cite{pekarek2001}-doped GaSe) are
also under investigation with the aim to find out new classes of
diluted magnetic semiconductors (DMS) of the form
A$_{1-x}^{III}$M$_{x}$B$^{VI}$, where A$^{III}$B$^{VI}$ is a III-VI
semiconductor and M is a transition metal ion. For instance, Mn has
been incorporated into $GaSe(0001)$ in samples grown from the melt,
and intriguing magnetic properties have been found \cite{pekarek}. A
short range anti-ferromagnetic ordering has been invoked to explain
the rather complex magnetic behavior, but a clear identification of
the short range coupling mechanisms related to these experimental
evidences is still missing. Moreover, a clear understanding of the
interplay between magnetism and electronic properties has not yet
been reported so far, mainly due to the difficulty of growing high
quality Mn-doped single crystals and control both the doping level
and possible phase segregations or the creation of defects and
vacancies upon doping. Also the local structure around Mn atoms at
the Mn:GaSe interface has not yet been probed, being the mechanism
of Mn diffusion in the lattice poorly investigated. This can be
important in order to relate the observed magnetic behavior to
either direct or superexchange interactions through $Mn-Mn$ or
$Mn-Se-Mn$ bonds, respectively.

Furthermore, recent studies \cite{lovejoy1, lovejoy2} on the
Mn:Ga$_{2}$Se$_{3}$ system have drawn the attention on this
interface, that is strictly related to the one we are currently
studying. The magnetic properties of the Mn:Ga$_{2}$Se$_{3}$ system
have also been reported \cite{dubinin2011}, suggesting weak
antiferromagnetic correlations in the bulk crystal.

Like the II-VI DMS, substitutional magnetic ions in the III-VI DMS
are found in a (distorted) tetrahedral environment. However, in
sharp contrast to the II-VI DMS, the III-VI semiconducting host
presents a two dimensional (2D) nature, at the origin of the renewed
interest in ultrathin layers of, e.g., GaSe and GaS \cite{003}. The
weak van der Waals bonding between the stacked four atom thick
layers (Se-Ga-Ga-Se) further enhances the two-dimensional nature of
this crystal. Because of its markedly nearly 2D structure, GaSe has
been considered in the past for angle-resolved photoemission (ARPES)
experiments, and a recent study has refocused the interest on this
aspect by providing high quality ARPES data supported by band
structure calculations of the bulk crystal\cite{plucinki}, while
electronic structure calculations and optical spectroscopy
experiments on few-layer GaSe sheets have been recently published
\cite{ryb}.

The present study is focussed on the electronic properties of the
Mn-GaSe interface obtained by evaporating Mn ions on the (0001)
surface of a ultra-high vacuum cleaved $\epsilon$-GaSe single
crystal. In the first part, we track the evolution of the Mn-GaSe
interface by evaporating at room temperature (RT) increasing
quantities of Mn on the GaSe surface. In this way we identify the
interface growth regimes, and in particular the balance between
cation diffusion through the surface and cation segregation at the
surface. Once the interplay between these processes was assessed, we
prepared a Mn:GaSe interface where the Ga$_{1-x}$Mn$_x$Se surface
alloying through cation diffusion is dominant over Mn segregation on
the surface, and we studied the electronic properties through
resonant photoemission and X-ray absorption spectroscopies. A
comparison is drawn with the Mn:CdTe interface, which displayed a
similar transition from cation diffusion to cation segregation
processes, as discussed in Ref.\cite{attenutaion}. This puts on a
solid ground the early speculations on the substitution of Ga by Mn
in the GaSe lattice, and provides an experimental evidence of the
capability of Mn to diffuse into the GaSe lattice. The similarity of
the present spectroscopic data with those found for
Cd$_{1-x}$Mn$_{x}$Te indicates that the diffusion process is rather
efficient across the surface of the GaSe system. Finally, a
characterization of the magnetic properties is reported, based on a
study the temperature dependence of the magnetization.

\section{Experimental details}
The GaSe single crystals have been grown by the Bridgman method
\cite{AGU}. The crystals were cleaved in ultra-high vacuum
conditions prior Mn evaporation. Mn layers were deposited at RT by
in-situ electron beam evaporation from an outgassed Mo crucible
loaded with metallic Mn flakes.  An Omicron EFM-3 triple evaporator
was used in all experiments. The deposition rate was properly
calibrated before evaporation on the GaSe cleaved surface. Three
interfaces have been produced during the experiments. (i) The first
was obtained by evaporating Mn with a constant flux of 500 nA
measured across the exit slit of the evaporator. No post-growth
annealing was carried out. The data collected from this interface
are reported in Section III.A. After an overall 180 second
deposition at this rate, the amount of deposited Mn was estimated to
be 2 ML. The second sample (Section III.B) was obtained by
evaporating 2.4 ML of Mn, and by annealing the interface at 400
$^{\circ}$C in ultra-high vacuum for 10 minutes to favor the Mn
diffusion process in the GaSe lattice. (iii) The third sample
(Section III.C-III.E) was produced at the BACH beamline, by
depositing a sub-ML of Mn at room temperature. As in the previous
case, a 10 minute annealing in ultra-high vacuum at 400 $^{\circ}$C
was carried out after the Mn evaporation.

X-ray absorption and resonant photoemission spectroscopy
measurements were performed at the BACH beam line of the Elettra
Synchrotron Light Source. The X-ray photoemission (XPS) data have
been collected at the Surface Science and Spectroscopy Lab of the
Universit\'{a} Cattolica (Brescia, Italy) with a non-monochromatized
dual-anode PsP x-ray source and a SCIENTA R3000 analyser, operating
in the transmission mode, which maximizes the transmittance and
works with a 30$^{\circ}$ acceptance angle. The stoichiometry of the
Mn:GaSe interfaces produced during the experiments was estimated by
measuring the peak area of the Mn, Ga or Se atomic species, properly
weighed by the photoemission cross sections and the analyzer
transmission. The surface sensitivity (XPS probe depth) at the
various kinetic energies (KE) was evaluated by Monte-Carlo
calculations of the depth distribution function with the algorithm
described in Ref.\cite{WernerDDF}, in order to include inelastic as
well as elastic electronic scattering, in the so-called transport
approximation\cite{JablonskiTA}. We define the calculated probe
depth for photoemission as the maximum depth from which 95\% of all
photo-emitted electrons can reach the surface. In case of
exponential attenuation of the signal, this would correspond to
three times the electron escape depth.

Magnetic measurements have been performed by means of a SQUID
magnetometer. A 5 kOe magnetic field has been applied parallel to
the sample plane to study the temperature dependence of the
magnetization in the range 2-360 K. A hysteresis cycle has been also
collected at room temperature with magnetic field ranging between
$\pm$10 kOe.

\section{Results and discussion}

\subsection{Cation diffusion vs. surface segregation}

\begin{figure}
\centering
\includegraphics[width=0.40\textwidth]{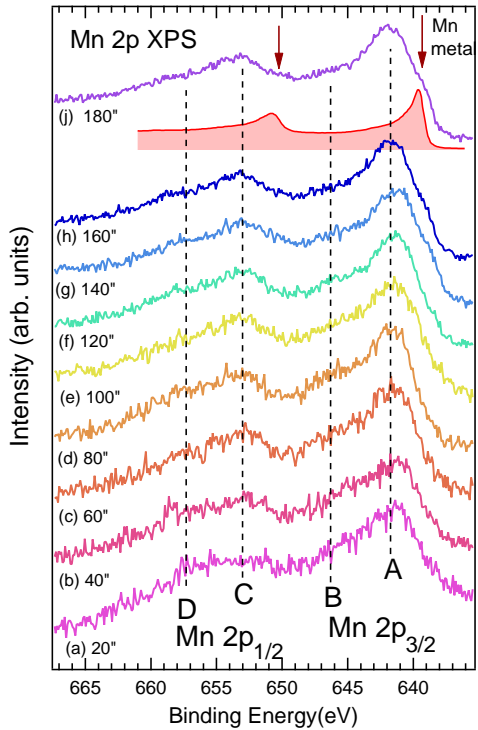}
\caption{(Color online) Sequence of Mn 2p spectra collected with Al
K$\alpha$ photon source for Mn deposited in steps (20 seconds for
each step) on $\epsilon$-GaSe(0001) surface. The time label
indicated the total Mn incremental deposition time at each step. The
spectra have been normalized to the peak height of the Mn 2p$_{3/2}$
component.} \label{mn series dep}
\end{figure}

In Figure \ref{mn series dep} a sequence of Mn 2p XPS spectra
collected after each Mn evaporation at RT on the freshly cleaved
$\epsilon$-GaSe surface is shown. As can be noticed, the Mn
2p$_{3/2}$ and Mn 2p$_{1/2}$ spin-split features of the Mn 2p core
line are detectable. Each spin-orbit split component is composed of
two broad peaks, denoted as A and B for the 2p$_{3/2}$ component,
and C and D for the 2p$_{1/2}$ component. These spectra are typical
of Mn in DMS, as will be discussed in the next Subsection.
Furthermore, the overall Mn 2p lineshape changes as the amount of
deposited Mn increases. In particular, starting from spectrum (f)
(120 seconds) a peak ascribed to metallic Mn is clearly detectable
on the low BE side of the Mn 2p main line, with a BE of 638 eV. The
Mn 2p XPS spectrum of a Mn thick film is also shown below spectrum
(j) (shaded area), to help identifying the contribution of metallic
Mn in the data.

\begin{figure}
\centering
\includegraphics[width=0.40\textwidth]{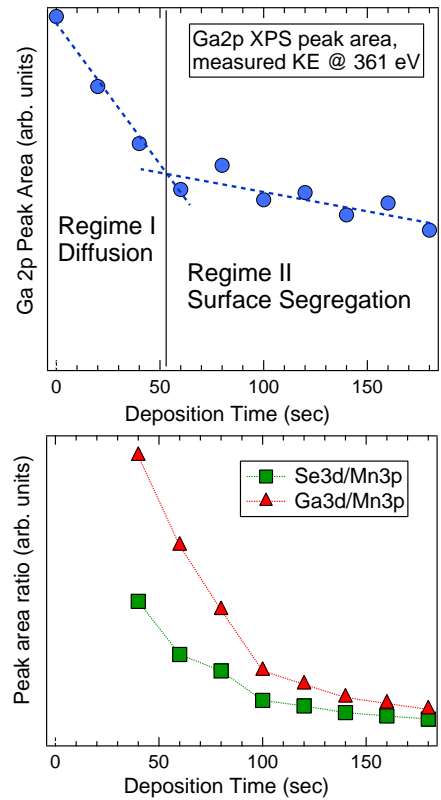}
\caption{(Color online) Top panel: Integrated intensities of Ga$2p$
core lines measured after each step of Mn deposition on
$\epsilon$-GaSe surface. Bottom panel: Se3d/Mn3p
    (filled squares) and Ga3d/Mn3p (filled triangles) peak area intensity
    ratios vs. deposition time.} \label{core_attenuation}
\end{figure}

In parallel with the Mn 2p core levels, also the Ga 2p core lines
have been collected. The integrated peak area of the Ga 2p is shown
in the top panel of Figure \ref{core_attenuation}. The Ga 2p signal
attenuation with deposition time is well detectable. The low kinetic
energy of the Ga 2p electrons makes the spectra more sensitive to
the surface layers. It is important to note that the Ga 2p
attenuation seems to follow two regimes. From 0 to about 50 seconds
the attenuation is steeper than that measured after 50 seconds. A
similar behavior was found for Mn deposited on CdTe single crystals,
and ascribed to Cd substitution with Mn \cite{attenutaion, sanga10}.
As in that case, we can ascribe the early steep decrease to Ga
cation substitution with Mn (corresponding to Cd cation substitution
with Mn in the CdTe host crystal), and the following slower decrease
with an overall screening of the Ga signal due to the growth of a Mn
overlayer on the crystal surface. Therefore, after a determined Mn
coverage, Mn diffusion to layers underneath the surface and Ga
substitution processes are hindered, resulting in the build-up of Mn
overlayers on the surface.

It should be noted that electrons from the Ga 2p core levels are
emitted with a kinetic energy of 361 eV. The XPS probe in GaSe for
these electrons is quite surface sensitive, as the calculated probe
depth is about 2.5 nm. Unfortunately, Se does not display core
levels with a comparable kinetic energy, as was the case of Cd 3d
and Te 3d \cite{attenutaion}, and it is not possible to evaluate
differences in the signal attenuation rate of Se surface sensitive
emission with respect to the Ga 2p case. To overcome this limit, a
set of shallow core levels (Se 3d, Mn 2p, and Ga 3d) with much
higher kinetic energies (about 1450 eV) has been collected. With
this KE, the probe depth is about 7.5 nm for the three shallow core
level emissions. On this basis, the Ga3d/Mn3p and Se3d/Mn3p ratio
vs. deposition time are shown in the bottom panel of
Fig.\ref{core_attenuation}. As can be observed, Ga and Se signal
have a steep decrease down to about 100 sec Mn deposition time, and
Ga decreases more rapidly than Se, suggesting that Ga is substituted
by Mn in the lattice. Above this limit both Ga 3d and Se 3d shallow
core levels display a much lower signal attenuation, indicating that
different mechanisms are at work to screen the photoemission signal,
very likely the prevalent growth of a Mn overlayer on the surface.
Consistently with the higher probe depth, the change of growth
regime here is found around 100 seconds rather than at about 60
seconds. We rationalize this finding by assuming that the
substitution of Ga with Mn is achieved at the early stages of
deposition for the topmost Ga layers (preferentially probed by the
Ga 2p attenuation), while the substitution mechanism in the
underlying layers remains active also at further deposition steps,
determined by the kinetics of Mn diffusion processes.

\begin{figure}
    \centering
    \includegraphics [width=0.5\textwidth]{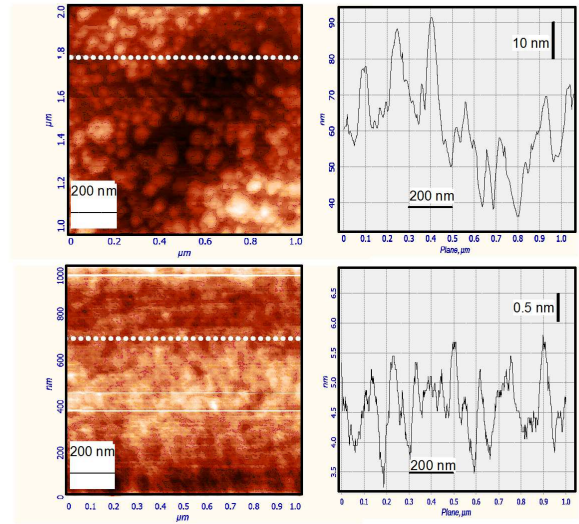}
    \caption{(Color online) AFM images (1$\mu m$ x 1$\mu m$ size) of the freshly
cleaved $\epsilon$-GaSe(0001) surface (bottom panel), and of the
Mn-GaSe(0001) interface after the last Mn evaporation stage at room
temperature (top panel).}
    \label{AFMfig}
\end{figure}

The build up of Mn on the surface leads to distinct features of the
sample surface, as shown in Fig.\ref{AFMfig}, top panel. In fact,
while the freshly cleaved surface shows a quite flat profile with
reduced roughness (Fig.\ref{AFMfig}, bottom panel), the surface
obtained after the last deposition stage is quite rough, with
round-shaped protrusions that could be related to Mn segregation on
the surface.

The Mn stoichiometry $x$ in the Ga$_{1-x}$Mn$_x$Se alloy was also
estimated by measuring the intensity of the Mn 2p core level peak
with respect to the Ga 3d and Se 3d core lines. After proper
normalization to photoemission cross-sections and analyzer
transmission, assuming a uniform distribution of Mn ions through the
topmost surface layers, the amount of Mn diffused into the crystal
after 60" evaporation (spectrum (c) of Fig.\ref{mn series dep},
roughly at the border between the two deposition regimes evidenced
in the bottom panel of Fig.\ref{mn series dep}) resulted to be
x=0.056$\pm$0.005 in Ga$_{1-x}$Mn$_x$Se.

\subsection{Core level photoemission of the Ga$_{1-x}$Mn$_x$Se alloy}

After the preliminary work so far described, we have been able to
identify the spectroscopic signatures of the conditions where the
alloying process, yielding Ga$_{1-x}$Mn$_x$Se, was dominant over the
Mn segregation on the surface. The Mn evaporation was repeated on
freshly cleaved GaSe surfaces and, after Mn evaporation, annealing
at 400 $^{\circ}$C in ultra high vacuum was also carried out to
further induce the alloying process through diffusion. The amount of
evaporated Mn (2.4 ML) exceeds the overall amount of Mn of Section
III.A by a factor 1.2, and a larger contribution of metallic Mn is
expected before annealing. However, the larger amount of Mn allowed
us to collect data with a a better statistics and obtain a good
reference Mn 2p XPS spectrum of the Mn:GaSe alloy after the
annealing treatment.

\begin{figure}
\centering
\includegraphics[width=0.49\textwidth]{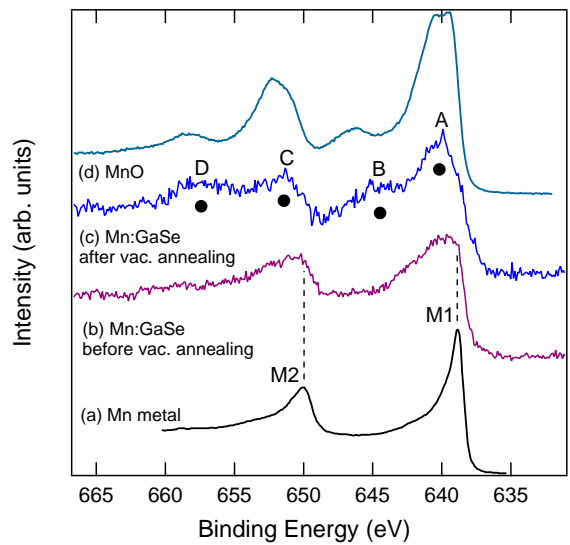}
\caption{(Color online) XPS spectra of the Mn 2p core lines
collected from (a) a thick Mn metallic film, (b) the Mn:GaSe
interface before UHV annealing, (c) Mn:GaSe interface after UHV
annealing (d) a MnO single crystal} \label{mn gs}
\end{figure}

Figure \ref{mn gs} shows the Mn 2p core line of the Mn:GaSe
interface prior and after annealing in vacuum. As observed, the four
spectral features of the vacuum annealed interface present strong
satellites (B and D) on the high BE side of the main lines (A and C)
of the Mn 2p spin-orbit split components.

We can exclude the presence of relevant oxygen contaminations, as
the measured Mn 2p XPS lineshape is quite different from that of MnO
(Figure \ref{mn gs}-d). Furthermore, we did not observe any signal
from oxygen within the sensitivity of our probe. The spectrum of the
as-deposited film (Figure \ref{mn gs}-b) shows two broad spin-orbit
split components, suggesting the presence of several contributions
that could be ascribed to both metallic Mn and Mn diluted in the
GaSe lattice. In fact, a comparison with the Mn 2p XPS core line
from metallic Mn (Figure \ref{mn gs}-a) indicates that the M1 and M2
features (marked by dashed vertical lines) can be ascribed to
metallic Mn. These features are progressively quenched with
annealing treatments (Figure \ref{mn gs}-c), indicating that the
annealing is quite effective to prevalently induce a substitution of
Ga by Mn atom, rather than a clustering of Mn on the GaSe(0001)
surface.

The peaks A and C are separated by the spin-orbit interaction and
the width of these two peak is ascribed to disorder effects, related
to replacement of Ga atom by Mn. On the high BE side of these peaks,
two satellites are also detectable (B and D), quite similar to those
found in Mn-based DMS, such as Cd$_{1-x}$Mn$_{x}$Te,
Zn$_{1-x}$Mn$_{x}$S and Ga$_{1-x}$Mn$_{x}$As \cite{sanga10, sanga11,
okabaya}. They are ascribed to charge transfer effects from the
ligand anions (Te, S or As, respectively) to the 3d levels of Mn
cations. These effects are usually accounted for in the frame of a
configuration interaction model where the electronic states involved
in the photoemission process are described by a linear combination
of several configurations (see, e.g. Ref \cite{okabaya} and Refs.
therein) such as 3d$^{n}$, 3d$^{n+1}$L, 3d$^{n+2}L^{2}$,where L
represents a hole in the ligand created by the charge transfer. The
ligand-to-3d charge-transfer energy is defined by
$\Delta$=E(d$^{n+1}$)-E(d$^{n}$). The intensity of B and D
satellites varies depending on the charge transfer energy $\Delta$,
as well as on the hybridization strength (T) between the p and d
orbitals involved in the charge transfer process (here from Se $4p$
to Mn $3d$). Therefore, the line-shape analysis of the Mn 2p core
levels shown in Figure \ref{mn gs} provides an evidence of Mn-Se
hybridization effects for the Mn:GaSe system. In Subsection D, a
detailed calculation of the Mn spectral weight in photoemission
through CI models will be carried out for the 3d levels in the
valence band region.

Finally, it is rather important to compare the present results with
those obtained on the Fe-GaSe interface\cite{iron1, iron2}. The
analysis of Fe 2p XPS lineshape in the Fe-GaSe interface does not
provide evidence of Fe-Se hybridization, being the Fe 2p XPS spectra
quite similar to that of metallic Fe, while Fe clustering effects
are found to be dominant. In turn, our measurements on the
electronic properties of the Mn:GaSe interface have shown the
capability of Mn to diffuse into the lattice with a remarkable
hybridizations with Se anions.

At this stage we cannot exclude that annealing in vacuum can also
trigger Mn desorption effects, but the results of the first
deposition (Section III.A) clearly indicate that alloying (i.e. Mn
diffusion though the lattice and Ga substitution with Mn without
resorting to any annealing) is the dominant process. According to
the GaSe layered crystal structure, the UHV cleaving should occur
between the two facing Se layers (i.e. through a breaking of the
Se-Se weak van der Waals bonds). The weakness of this bond should
also favor the diffusion of Mn through the lattice between the Se
layers, and eventually the Mn hybridization with Se.

\subsection{X-ray absorption from the Ga$_{1-x}$Mn$_x$Se alloy}

\begin{figure}
\centering
\includegraphics[width=0.45\textwidth]{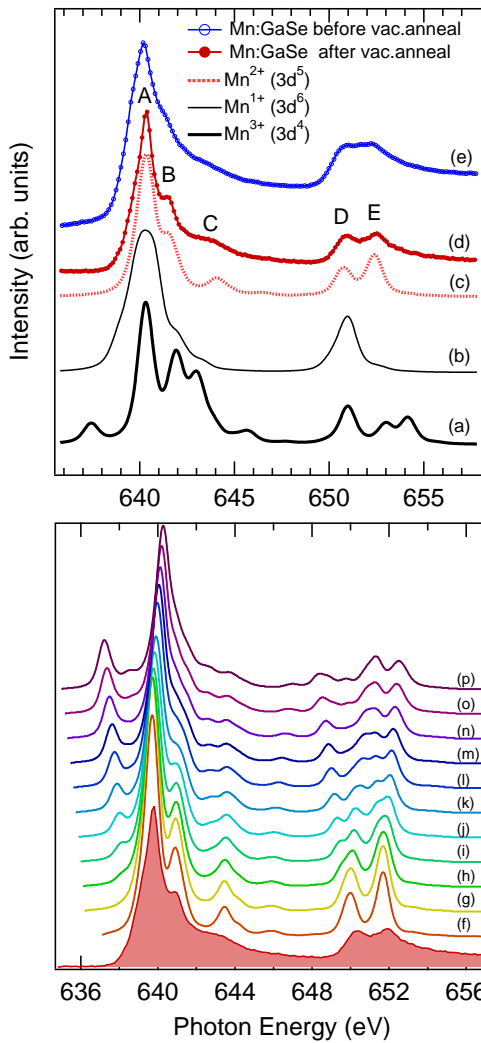}
\caption{(Color online) Top panel. Experimental Mn L-edge XAS
spectra of Mn deposited on GaSe surface before annealing figure (e),
after annealing figure (d). Calculated atomic Mn L-edge XAS spectra
(zero crystal field) for the +3 oxidation state (3d$^{4}$) (a), +1
oxidation state (3d$^{6}$) (b), and $+2$ oxidation state (3d$^{5}$)
(c). Bottom panel. Comparison between the experimental XAS spectrum
of the annealed Mn:GaSe interface (shaded area) and the calculated
XAS spectrum for the atomic Mn$^{2+}$ 3d$^{5}$ configuration with a
tetrahedral crystal field $10 Dq$ ranging from 0.0 eV to 2.5 eV}
\label{xascfmgs}
\end{figure}

The Mn L-edge XAS spectra are shown in Figure \ref{xascfmgs}. In
particular, the data obtained after the Mn deposition (e) and after
annealing at 400$^{o}$C and collected at RT (d) are presented. The
as-deposited Mn-doped GaSe (e)  shows before annealing the presence
of both metallic and a reacted Mn-GaSe interface, as appearing from
an overall smooth XAS lineshape, with minor modulations that will
ultimately evolve into the post-annealing XAS lineshape (d). Indeed,
after annealing at 400$^{o}$C, sharper features (labeled as A, B, C,
D, and E) appear, and the comparison with multiplet calculations for
a Mn$^{2+}$ 2p$^{6}$3d$^{5}$ $\rightarrow$ 2p$^{5}$3d$^{6}$
electric-dipole allowed transition (c) unambiguously shows that the
measured spectrum can be ascribed to a Mn$^{2+}$ ion in the GaSe
matrix. Similar transition calculated for Mn$^{1+}$ (b) and
Mn$^{3+}$ (a) ions do not fit the experimental data.

The remarkable similarity with the XAS spectrum predicted for the
Mn$^{2+}$ calculation is particularly helpful for the interpretation
of the electron spectroscopy results. This will justify the
assumption at the basis of CI calculations for the Mn 3d spectral
weight in the valence band (see next Section) where an Mn$^{2+}$ ion
will be assumed as the ionic configuration in the parameterized CI
model.

In the bottom panel of Figure \ref{xascfmgs}, we have shown a set of
calculated Mn XAS spectra in a tetrahedral T$_{d}$ symmetry,
starting from zero crystal field to 10Dq=2.5 eV, where 10Dq is the
crystal field splitting of the 3d orbitals. It is important to note
that crystal field effects seem to be rather limited up to 10Dq=0.75
eV. At this energy, two features appear on the low photon energy
sides of the calculated Mn L$_{III}$ and Mn L$_{II}$ edges, that
have no counterpart in the experimental data. Therefore, we assume
that crystal field effects are negligible. This remark will also be
at the basis of the CI model for the valence band calculations,
where crystal field splitting will be set to zero.

\subsection{Valence band resonant photoemission at the Mn L-edge}

\begin{figure}
\centering
\includegraphics[width=0.45\textwidth]{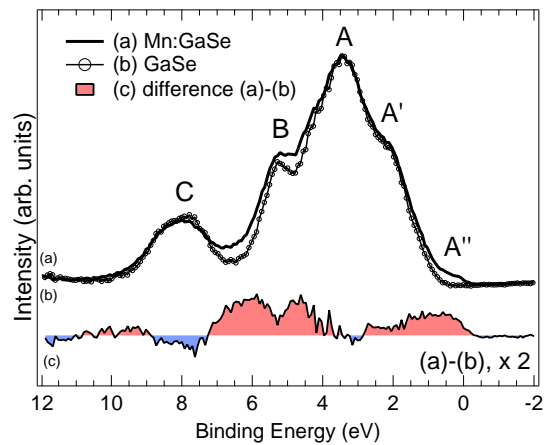}
\caption{(Color online) Valence band spectra of the clean GaSe (b)
and of the Mn-doped, annealed, GaSe single crystal (a). Difference
between the spectra of the doped and the clean system  (c). }
\label{Vbm1}
\end{figure}

The valence band spectra of the clean GaSe and of the Mn-doped,
annealed, GaSe single crystals are shown in Figure \ref{Vbm1} (b)
and (a), respectively. The photoemission spectra have been collected
with a photon energy of 797 eV and have been normalized to the
maximum of the valence band emission (peak A). Both spectra show a
main line with three features labeled A', A, and B, and a peak C at
higher binding energies. When Mn is evaporated on the GaSe cleaved
surface, the main changes that can be observed are the appearance of
a feature A" at the Fermi edge, and an increase of the spectral
weight in the regions between the peaks A and B and the peaks B and
C. The curve displayed in Fig \ref{Vbm1} (c) represents the
difference between the spectra (a) and (b). This difference confirms
the increase of spectral weight upon Mn deposition in the 3-7 eV
binding energy range, and in the region just below the Fermi edge
(BE=0-2 eV), while a decrease of the intensity is found below peak C
after Mn deposition and annealing at 400 $^{\circ}$C.

The main features of the present experimental data can be
interpreted on the basis of the calculated DOS so far published
\cite{plucinki, DOS2, DOS3}. Indeed, the observed experimental peaks
have a counterpart in, e.g., the DOS calculated by Plucinki \emph{et
al.} \cite{plucinki} that identify three regions (I, II, and III) in
the valence band (see Fig.3 of Ref.\cite {plucinki}). Region I
corresponds to the observed peaks A and A', region II to peak B and
region III to peak C. From the analysis of the projected DOS on Se
s,p states and Ga s,p states, it is rather interesting to observe
that Ga mainly contributes in the region below peak C, i.e. in the
region where a spectral weight decrease is observed upon Mn doping
and annealing. This is in agreement with the assumption of Ga
substitution with Mn atoms, as remarked in Subsection A. Finally,
the states appearing at the Fermi edge (A'') can be ascribed to
unreacted Mn at the surface.

\begin{figure}
\centering
\includegraphics[width=0.45\textwidth]{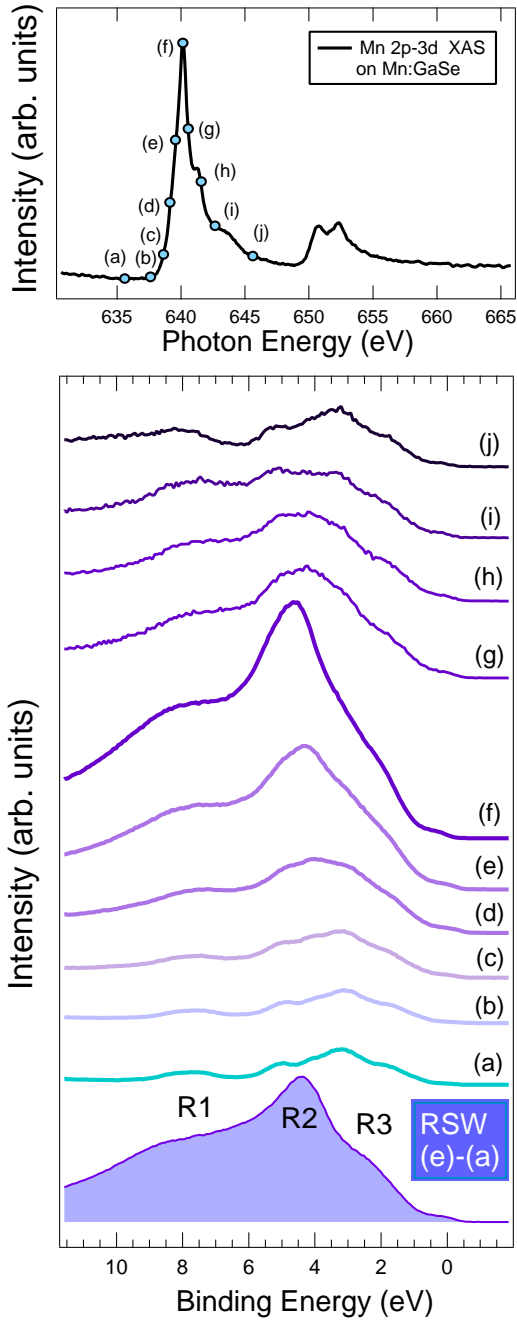}
\caption{(Color online) Top panel: XAS spectrum of the Mn-GaSe
interface collected at the Mn 2p-3d edge . The dots indicate the
photon energies chosen to collect the, resonant, valence band
photoemission spectra. Botton panel: Set of valence band spectra
collected at different photon energies across the Mn 2p-3d edge.
Resonant spectral weight (RSW, \textbf{x 1.2}) obtained from the
difference between the off-resonance spectrum (a) and resonant VB
spectrum (e) collected at 635.5 eV and 639.5 eV, respectively.
Difference between the resonant and off-resonance spectra.}
\label{rgs}
\end{figure}

In order to enhance the Mn contribution to valence band states, a
ResPES study at the Mn 2p-3d absorption edge has been carried out.
The results are shown in Fig \ref{rgs}. As first, on the XAS
spectrum (top panel) the photon energies (a to j) selected to
collect ResPES data are indicated. The whole set of ResPES data is
shown in the bottom panel. The data span a photon range across the
Mn L$_{III}$ threshold. The VB spectra show a clear enhancement of
the spectral weight with a photon energy of about 640 eV. At this
energy a peak around BE= 4.5 eV in the valence band shows a
remarkable intensity enhancement. The difference between the
resonant (e) and off-resonance (a) spectra (hereafter denoted as
resonating spectral weight, RSW) is shown in panel C. Here it is
clearly seen that the RSW is determined by a peak at 4.5 eV (R2) and
by two broad features at about BE= 6-8 eV (R1) and BE= 1-3 eV (R3).

We have used spectrum (e) rather than (f), as it is known
\cite{mnge2012} that for spectra collected with photon energies at
the maximum of the absorption threshold [spectrum (f) in the present
case] the weight of the Auger emission channel is not negligible and
the maximum of the valence band emission is already shifted to
higher BE with respect to spectrum (e), where the resonant Raman
Auger channel (RRAS) is still dominant.

It is worth observing that the large R2 resonance is found in the
energy range where the difference spectrum of Fig.\ref{Vbm1}-c shows
the largest difference between the doped and undoped surfaces,
confirming that a large Mn contribution should be expected at least
in this energy range.

Further insight into the origin of the three spectral features (R1,
R2, and R3) in the RSW can be obtained at the light of parameterized
CI calculations for the valence band.

\begin{figure}
\centering
\includegraphics[width=0.46\textwidth]{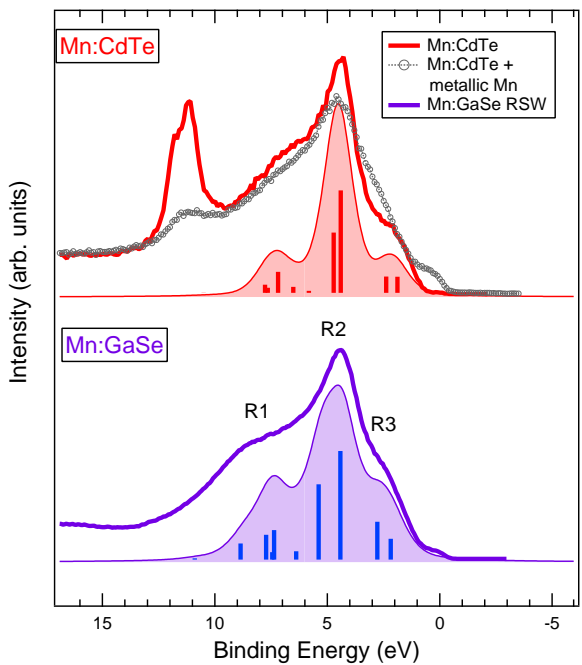}
\caption{(Color online) Experimental data and CI calculations for
the Mn:CdTe (top panel, adapted form Ref. \cite{attenutaion}) and
the Mn:GaSe (bottom panel) interfaces. The shaded areas represent
the CI calculations, while the vertical bars indicate the
eigenenergies obtained by solving the Hamiltonian matrix. The height
of each bar is proportional to the square of the projection of the
eigenvector on the ground, initial, state. All energies are given in
eV.} \label{cigs}
\end{figure}

Impurity-cluster CI calculations of the Mn 3d spectral weight are
shown in Fig.\ref{cigs}. In the CI approach several configurations,
denoted as $d\,^{n}$, $d\,^{n+1}\, \underline{L}$ ($\underline{L}$
denotes a ligand hole, here a hole on Te 4p orbitals) are used to
describe the open shell of the 3d transition metal ion during the
photoemission process. The spectral weight in a photoemission
experiment is calculated, in the sudden approximation, by projecting
the final state configurations ($|\Psi_{i,fs}\rangle$) on the the
ground state, i.e.
$I_{XPS}(BE)\propto\sum_{i}\left|\langle\Psi_{GS}|\Psi_{i,fs}\rangle
\right|^{2}\delta(BE-\varepsilon_{i})$
where $|\Psi_{GS}\rangle$ represents the ground state (GS)
wavefunction, and the sum is run over all final state configurations
$|\Psi_{i,fs}\rangle$ with energy $\varepsilon_{i}$. Where required,
proper fractional parentage coefficients can be used, as was done in
the present case.

The calculations have been carried out following the scheme
presented by Fujimori et al. \cite{mnconfi} for several Mn doped
semiconductors, based on a 3d$^{5}$ initial state of the transition
metal atom \cite{fuji}. The results are shown in Fig.\ref{cigs},
along with those obtained on a Mn-doped CdTe single crystal
\cite{attenutaion}. The parameter set used in the calculation is
reported in Table \ref{cimn}.

The comparison with the RSW detected under the same conditions for
the Mn-doped CdTe crystal is rather interesting. The gray spectrum
(empty circles) in the top panel represents the resonance spectrum
collected from an heavily doped CdTe single crystal (adapted form
Ref. \cite{attenutaion}), which shows a contribution at the Fermi
level of metallic Mn states similar to the states observed in the
GaSe host crystal, though these states are less intense in the GaSe
host system. Also in the CdTe case, three peaks appear in the
resonant spectrum, but the relative weight and width of these peaks
are different from the Mn:GaSe case. Peak R2 is larger in the GaSe
host, and the separation between peak R1 and R2 is larger in the
GaSe host with respect to the CdTe case. These differences have been
considered as constraints in the calculations. In particular, the
calculations for the GaSe case have been obtained by setting the
crystal field to zero and by considering a larger charge transfer
energy (2.95 eV) as compared to the CdTe host (2.0 eV, Table
\ref{cimn}). The first assumption is justified by the lack of
relevant crystal field effects observed in XAS, whereas the second
is justified by the larger band gap of GaSe with respect to CdTe, as
the charge transfer energy is usually assumed to scale with the
energy gap of the host crystal (E$_g$=1.49 eV in CdTe, E$_g$=2.02 eV
in GaSe). This choice of the parameter set resulted in a broadening
of the calculated peak below R2 and in the intensity increase and BE
shift of the calculated spectral weight below R1. Also the
calculated spectral weight below R3 is increased, in agreement with
the measured data. We used the free ion Racah parameter for
Mn$^{2+}$, B=0.126, C=0.421, and the crystal field was set at 0.4 eV
for the CdTe host and 0 eV for the GaSe host.

\begin{table}

\centering \caption{Parameter values used for the parameterized CI
calculation of the Mn 3d spectral weight. The following values are
specified: the charge transfer energy $\Delta$, the d-d correlation
energy U, and the pd$\sigma$ hybridization integral}

\label{tab222}

\begin{tabular}[c]{ccccccccc}
\hline
 \hline
\hline    Host     & $\Delta$      &   U           &   pd$\sigma$     &  \\
   $ $    Matrix        & (eV)          & (eV)          &   (eV)    &  \\
\hline     GaSe      & 2.95          & 6.4           & -1.25     &   \\
\hline     CdTe      & 2             & 5             & -1.1      &   \\

\hline
\end{tabular}
\label{cimn}
\end{table}

\subsection{Magnetic properties}
\begin{figure}
\centering
\includegraphics[width=0.45\textwidth]{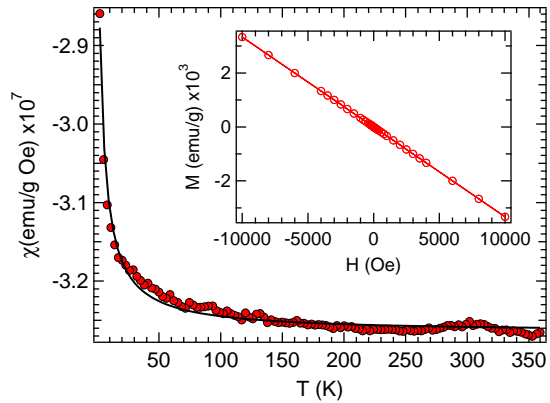}
\caption{(Color online) Magnetic susceptibility vs. temperature
curve for the Mn-doped GaSe sample. Inset: magnetization M vs.
external applied field H measured at 300 K.} \label{magnetic}
\end{figure}

Figure \ref{magnetic} shows the temperature dependence of the
magnetic susceptibility $\chi$, evaluated with respect to the
unit-mass of the considered sample. The overall susceptibility is
always negative disclosing a dominant diamagnetic behavior. This is
confirmed by the negative slope of the M vs. H curve measured at
room temperature, reported in the inset on Fig. \ref{magnetic}. The
negative dominant background is not surprising as the GaSe host is
diamagnetic and the Mn:GaSe interface represents a small fraction of
the sample (few nanometers with respect to the overall single
crystal thickness, that is about 100 microns). However, though
negative,  the magnetic susceptibility curve reveals a signature of
paramagnetism, with a decrease of values with temperature that is
fit by the curve representing the Curie-Weiss law (continuous line
in Fig. \ref{magnetic}), with very small Weiss constant value (about
-3 K).

This effect can be ascribed to the topmost layers of the sample,
i.e. those hosting the Mn dopant after the temperature induced
diffusion. This behavior is quite different from that observed in
bulk Mn:GaSe single crystals \cite{pekarek}, which show a complex
behavior with a quenching of magnetization at low temperatures,
i.e., where we observe the steady increase of paramagnetic signal.
The magnetic behavior is also different in many details from that
observed for other known bulk phases that can be regarded as
possible segregated phases in the growth of the Mn:GeSe interface,
such as MnSe and MnSe$_2$ \cite{88, 99}, and MnGa$_2$Se$_4$
\cite{89}. In turn, closer similarities, i.e. a virtually
paramagnetic behavior, are found with respect to the case of
monoclinic crystals of (Ga$_{1-x}$Mn$_{x}$)$_2$Se$_3$ discussed in
Ref. \cite{dubinin2011}.

\section{Conclusions}
We have been able to prepare well characterized Mn:GaSe interfaces,
with evidence of the formation of a Ga$_{1-x}$Mn$_{x}$Se alloy below
the GaSe surface. Alloying is obtained already after evaporation at
room temperature. The Mn deposition in the present study spanned two
regimes. In the early regime the Mn diffusion through the surface
was the dominant mechanism, while in the second regime the
segregation of Mn layers on the GaSe surface was the most likely
process. Unlike Fe-GaSe interfaces, where iron clustering effects
are dominant and no trace of Fe-Se hybridization is found upon the
analysis of Fe 2p XPS lineshape \cite{iron1, iron2}, our
measurements on the electronic properties of the Mn:GaSe interface
have shown the capability of Mn to diffuse into the lattice with a
remarkable hybridizations with Se anions. Magnetic measurements
evidence a paramagnetic behavior for the Mn-doped interface, while
the dominant behavior is diamagnetic, due to the bulk of the GaSe
host crystal.

\end{document}